\documentclass[3p]{elsarticle}
\usepackage{graphicx}
\usepackage{dcolumn}
\usepackage{xcolor}
\usepackage{bm}
\usepackage{array,multirow}
\usepackage{amssymb}
\usepackage{amsmath}
\usepackage{ulem}
\usepackage{lineno,hyperref}
\usepackage{makecell}
\modulolinenumbers[5]
\usepackage{comment}
\usepackage{soul}

\begin{document}

\begin{frontmatter}
\title{Geometric scaling of elastic  $pp$ cross section at the LHC}
\author{Micha{\l} Prasza{\l}owicz\fnref{mpfootnote}}
\address{Institute of Theoretical Physics, Faculty of Physics, Astronomy and Applied Computer Science, \\
Jagiellonian University, S. {\L}ojasiewicza 11,
30-348 Krak{\'o}w, Poland.}
\fntext[mpfootnote]{michal.praszalowicz@uj.edu.pl}

\begin{abstract}
We show that geometric scaling conjectured and observed at
the ISR more than 50 years ago, still holds at the LHC. We discuss
regularities of the dip\,-\,bump structures of the differential elastic cross sections, emphasizing the fact that the ratio of bump to dip positions
is constant from the ISR to the LHC. Applying crossing and analyticity we
identify imaginary and real parts of the scattering amplitude and compute
the $\rho$ parameter and the ratio of bump to dip values of the differential $pp$ cross sections. We also discuss the energy dependence
of the total elastic cross section and the violation of geometrical scaling outside the dip\,-\,bump region at the LHC.
\end{abstract}


\end{frontmatter}

\section{Introduction}

In this paper we argue that geometric scaling conjectured and observed at the
ISR more than 50 years ago \cite{DiasDeDeus:1973lde,Buras:1973km,Barger:1974vg}, still holds
at the LHC. This is not so obvious, since recent attempts to find a scaling
law for elastic $pp$ cross sections at the LHC energies have pointed to a
different scaling \cite{Baldenegro:2022xrj,Baldenegro:2024vgg}, possibly
associated with high energy saturation~\cite{Peschanski:2024tlr}. For other
studies on scaling properties of elastic cross sections see
\cite{Dremin:2012qd, Csorgo:2019ewn}.

There exist three sets of total and two of differential $pp$ high energy data.
The ISR data \cite{Nagy:1978iw,Amaldi:1979kd} collected at CERN in 1970's
cover energies $20~\mathrm{GeV}\lesssim W=\sqrt{s}\lesssim60~\mathrm{GeV}$,
recent LHC data by the TOTEM
\cite{TOTEM:2011vxg,TOTEM:2015oop,TOTEM:2017sdy,TOTEM:2018psk} and ATLAS
\cite{ATLAS:2014vxr,ATLAS:2016ikn,ATLAS:2022mgx} Collaborations at
$3~\mathrm{TeV}\lesssim W=\sqrt{s}\lesssim13~\mathrm{TeV}$ and the cosmic rays
total cross section data, which, however, have rather large errors (see
Fig.~\ref{fig:total}).

Starting from the ISR energies total $pp$ cross sections rise with energy, but
the increase is faster at the LHC than at the lower energies. To show this in
Ref.~\cite{Baldenegro:2024vgg}  a power law fit was performed in the
restricted energy ranges, separately for the ISR data \cite{Amaldi:1979kd}  and for the LHC
data~\cite{Nemes:2019nvj}. The results
are shown in Table~\ref{tab:sigmas}. One can see that at the  ISR all $pp$
cross sections (total, elastic and inelastic) increase with energy with  a
universal power of $\simeq0.11$, which
  led to the formulation of geometric scaling (GS)  in the 1970's \cite{DiasDeDeus:1973lde,Buras:1973km,Barger:1974vg}. 
  This is no longer true at the LHC, casting
 doubt on the possibility that  GS still holds at the LHC~\cite{Baldenegro:2022xrj,Baldenegro:2024vgg}.

\renewcommand{\arraystretch}{1.5} \begin{table}[h]
\centering
\begin{tabular}
[c]{|c|c|c|c|c|}\hline
& elastic & inelastic & total & $\rho$\\\hline
ISR & $W^{0.1142\pm0.0034}$ & $W^{0.1099\pm0.0012}$ & $W^{0.1098\pm0.0012}$ &
$0.02-0.095$\\\hline
LHC & $W^{0.2279\pm0.0228}$ & $W^{0.1465\pm0.0133}$ & $W^{0.1729\pm0.0163}$ &
$0.15-0.10$\\\hline
\end{tabular}
\caption{Energy dependence of the integrated $pp$ cross sections for the
energies $W=\sqrt{s}$ at the ISR \cite{Amaldi:1979kd} and at the LHC
\cite{Nemes:2019nvj}, and the $\rho$ parameter
\cite{Amaldi:1979kd,TOTEM:2017asr}. Table from Ref.~\cite{Baldenegro:2024vgg}%
.}%
\label{tab:sigmas}%
\end{table}
\renewcommand{\arraystretch}{1.0}

GS for the integrated cross sections can be best visualized in the impact parameter space.
To see this, let's consider the textbook formulas for the total and elastic cross
sections  in the normalization, where the scattering
amplitude is dimensionless
\begin{align}
\sigma_{\text{tot}}(s) &  =2%
{\displaystyle\int}
d^{2}\boldsymbol{b}\,\operatorname{Im}T_{\text{el}}(s,b),\nonumber\\
\sigma_{\text{el}}(s) &  =%
{\displaystyle\int}
d^{2}\boldsymbol{b}\,\left\vert T_{\text{el}}(s,b)\right\vert ^{2}%
.\label{eq:sigmasb}%
\end{align}
Very often one uses eikonal parametrization  $\operatorname{Im}T_{\text{el}}(s,b)=1-\exp(-\Omega(s,b))$
where  $\Omega$  is called opacity.
Inelastic cross-section is given as $\sigma_{\text{inel}}(s)=\sigma
_{\text{tot}}(s)-\sigma_{\text{el}}(s)$.

GS in impact parameter space means that $\operatorname{Im}%
T_{\text{el}}(s,b)=\operatorname{Im}T_{\text{el}}(b/R(s))$, where $R(s)$ is a
so called interaction radius to be determined from the data~\cite{DiasDeDeus:1973lde}. It follows that the
amplitude, which is formally a function of two independent variables $b$ and
$s$, depends in fact only on one dimensionless scaled variable $B(s)=b/R(s)$.
Changing variables in the first equation (\ref{eq:sigmasb}) $b\rightarrow
B=b/R(s)$ one obtains%
\begin{equation}
\sigma_{\text{tot}}(s)=2R^{2}(s)%
{\displaystyle\int}
d^{2}\boldsymbol{B}\,\operatorname{Im}T_{\text{el}}(B).\label{eq:sigtotscaled}%
\end{equation}
If the real part of the elastic scattering amplitude
can be neglected, as suggested by the
smallness of the parameter $\rho=\operatorname{Re} {T}_{\text{el}%
}(s,t=0)/\operatorname{Im}{T}_{\text{el}}(s,t=0)$ in momentum space, see
Table~\ref{tab:sigmas}, we get that also $\sigma_{\text{el}}(s)\sim R^{2}(s)$
and as a consequence also $\sigma_{\text{inel}}(s)\sim R^{2}(s)$ \cite{Barger:1974vg}. As mentioned
above GS of integrated  cross sections is clearly seen at the ISR, but not at the LHC.

\section{Elastic differential cross section and the dip\,-\,bump structure}

Let us now recall main features of $pp$ differential elastic cross section
$d\sigma_{\mathrm{el}}/dt$. One observes a rapid decrease for small $|t|$,
then a minimum at $t_{\mathrm{dip}}$, followed by a broad maximum at
$t_{\mathrm{bump}}$. This dip\,-\,bump structure emerges, as we will see later, from the Fourier
transform of the impact parameter amplitude~\cite{DiasdeDeus:1975ybq} and is related to the zeros and extrema of $J_0$ Bessel function.
A universal and striking property was reported in
Ref.~\cite{Baldenegro:2024vgg}, namely that the ratio of bump to dip
{\sl positions} is constant over the large energy span from the ISR to the
LHC
\begin{equation}
\mathcal{T}_{\mathrm{bd}}(s)=|t_{\mathrm{bump}}|/|t_{\mathrm{dip}}|
=1.355\pm0.011\, . \label{eq:Tbd}%
\end{equation}
On the contrary, the ratio of the cross section {\sl values}
\begin{equation}
\mathcal{R}_{\mathrm{bd}}(s)=\frac{ d\sigma_{\mathrm{el}}
/d|t|_{\mathrm{{bump}}}} { d\sigma_{\mathrm{el}} /d|t|_{\mathrm{dip}}}\, ,
\label{eq:Rbd}%
\end{equation}
which saturates at the LHC energies at a value of approximately 1.8, is rather
strongly energy dependent at the ISR (see e.g. Fig.~2 in
Ref.~\cite{TOTEM:2020zzr} and Fig~\ref{fig:Rbd}).

Property (\ref{eq:Tbd}) implies that dip and bump positions should be
identical at all energies if $d\sigma_{\mathrm{el}}/dt$ is plotted in terms of
a scaling variable $\tau=f(s) |t|$. In order to find $f(s)$ at the LHC
Ref.~\cite{Baldenegro:2024vgg} performed a power law fit to the dips
$|t_{\mathrm{dip}}|= \mathrm{const}\times(W/(1~\mathrm{TeV}))^{-\beta}$ with
the result $\beta=0.1686\pm0.0037$ and $\mathrm{const}=0.732\pm0.003$%
~GeV$^{2}$. Therefore $f(s)\sim W^{\beta}$. Bumps are reproduced due to the
property (\ref{eq:Tbd}). Note that, within experimental accuracy, $\beta$ is
equal to exponent of the total cross section (see Table~\ref{tab:sigmas}).
Hence from Eq.~(\ref{eq:sigtotscaled}) we can take $f(s)=R^{2}(s)\sim
\sigma_{\mathrm{tot}}(s)$.

At the ISR the situation is very similar. The authors of
Ref.~\cite{Buras:1973km} used $R^{2}(s)\sim\sigma_{\mathrm{inel}}(s)$,
however, since the energy dependence of $\sigma_{\mathrm{inel}}$ and
$\sigma_{\mathrm{tot}}$ is the same, they could have used $f(s)=R^{2}%
(s)\sim\sigma_{\mathrm{tot}}(s)$ as well, which they did in Ref.~\cite{DiasdeDeus:1977af}.

In order to further discuss GS it is more convenient to work in the momentum
representation%
\begin{equation}
T_{\text{el}}(s,b)=%
{\displaystyle\int}
\frac{d^{2}\boldsymbol{q}}{(2\pi)^{2}}e^{i\,\boldsymbol{bq}}T_{\text{el}%
}(s,t).\label{eq:Fourier}%
\end{equation}
Here $b=\left\vert \boldsymbol{b}\right\vert $, $t=- \boldsymbol{q}^{2}$.
Then, from Eq.~(\ref{eq:sigmasb})%
\begin{align}
s \sigma_{\text{tot}}(s)  &  =2\operatorname{Im}\tilde{T}_{\text{el}%
}(s,0).\label{eq:sigtotc}%
\end{align}
Following Refs.~\cite{DiasDeDeus:1973lde,Buras:1973km} we prefer work in a
normalization where the elastic amplitude is dimensionless $\tilde
{T}_{\text{el}}(s,t)=s {T}_{\text{el}}(s,t)$. In this normalization the total
elastic cross-section reads%
\begin{align}
\sigma_{\text{el}}(s)  &  =\frac{1}{4\pi s^{2}}%
{\displaystyle\int}
dt\left\vert \tilde{T}_{\text{el}}(s,t)\right\vert ^{2}.\label{eq:dseldt}%
\end{align}

Now, we would like to parametrize $\tilde{T}_{\text{el}}(s,t)$ in a way which
explicitly obeys GS and reproduces the energy dependence of the total cross
section (\ref{eq:sigtotc}),  neglecting for the moment the real part. In
momentum representation GS means that $\tilde{T}_{\text{el}}(s,t) \rightarrow \tilde
{T}_{\text{el}}(s,\tau)$,\footnote{Dependence of $\tilde{T}_{\text{el}}(s,\tau)$ on $s$ is only
in the normalization factor.}
where $\tau=\left\vert t\right\vert R^{2}(s)$. Then,
to recover the energy dependence $\sigma_{\text{tot}}(s) \sim R^{2}(s)$ in
Eq.~(\ref{eq:sigtotc}), we assume that (we will be more precise later)%
\begin{equation}
\tilde{T}_{\text{el}}(s,\tau) = i s R^{2}(s)\Phi(\tau) \, .\label{eq:ansatz0}%
\end{equation}
Here $\Phi(\tau)$ is a dimensionless real function of the scaling variable
$\tau$.

Substituting (\ref{eq:ansatz0}) to Eq.~(\ref{eq:dseldt}) we get
\begin{align}
\sigma_{\text{el}}(s)  &  =\frac{R^{2}(s)}{4\pi}{\displaystyle\int} d\tau\,
\Phi^{2}(\tau).\label{eq:dseldtau}%
\end{align}
We see that the energy dependence of $\sigma_{\text{el}}(s)$ is indeed, as at
the ISR, the same as $\sigma_{\text{tot}}(s)$. We will discuss later why this
relation may be broken at the LHC.

To proceed further, let us observe that function $\Phi(\tau)$ must have at
least one zero~\cite{DiasdeDeus:1975ybq,DiasdeDeus:1977af}. Indeed, taking an inverse Fourier transform of
(\ref{eq:Fourier}) we obtain
\begin{equation}
\Phi(\tau)=2\pi\int dB\, B\, J_{0}(B \sqrt{\tau})\, \operatorname{Im}
T_{\mathrm{el}}(B) \, .
\label{eq:PhiHD}
\end{equation}
Since $\operatorname{Im} T_{\mathrm{el}}(B)$ is typically a smooth localized
function (for example in the black disc toy model $\operatorname{Im}
T_{\mathrm{el}}(B) \sim\Theta(1-B)$), function $\Phi(\tau)$ has zeros
corresponding to the zeros of the Bessel function $J_{0}$. Notably, the first
zero corresponds to the dip of the amplitude square~\cite{DiasdeDeus:1975ybq,DiasdeDeus:1977af}. At this point we cannot
neglect the real part. Further, the minimum of the amplitude after the first
zero corresponds to the bump in the cross section.

\begin{figure}[h]
\centering
\includegraphics[height=6.5cm]{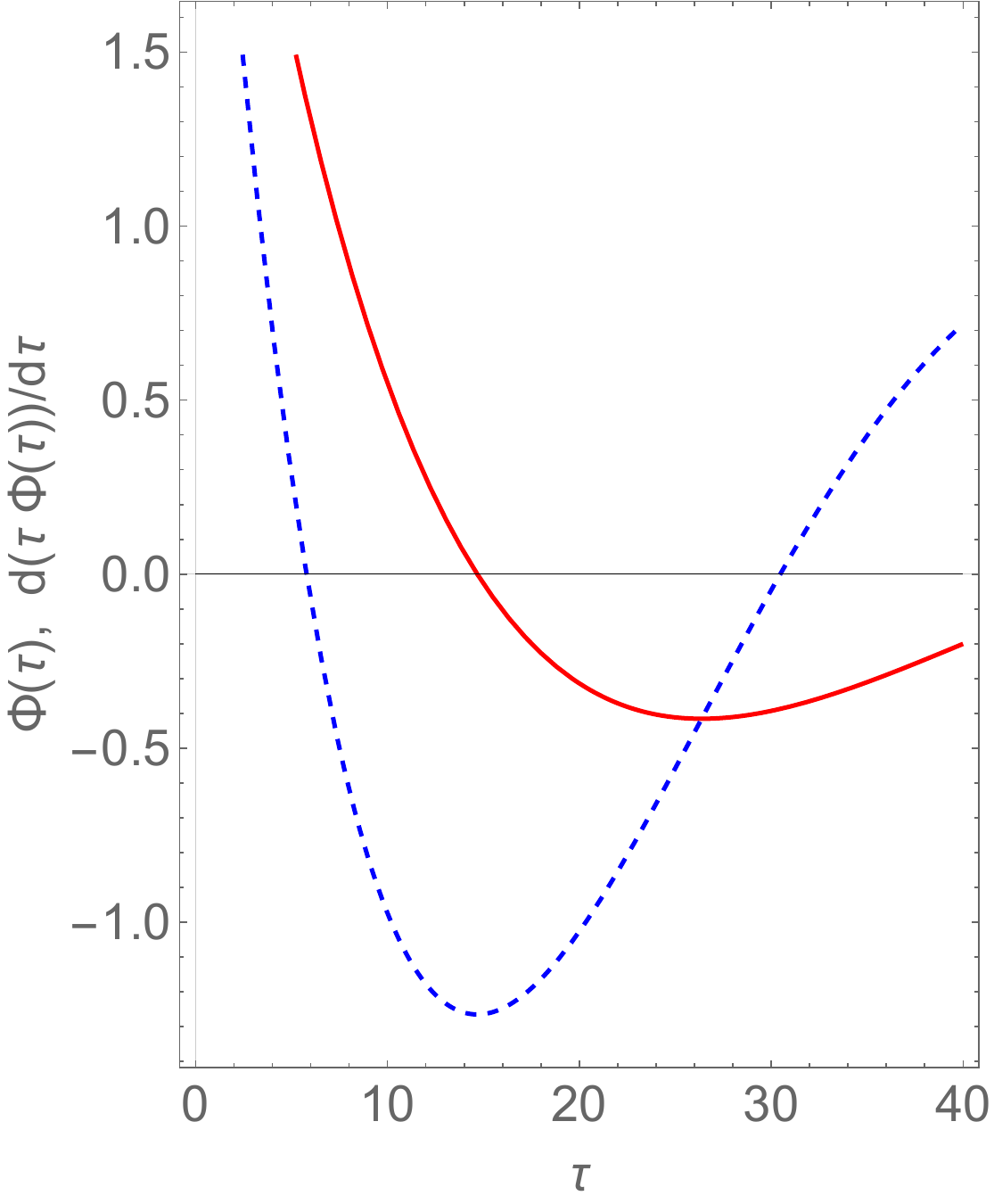} ~~ \includegraphics[height=6.65cm]{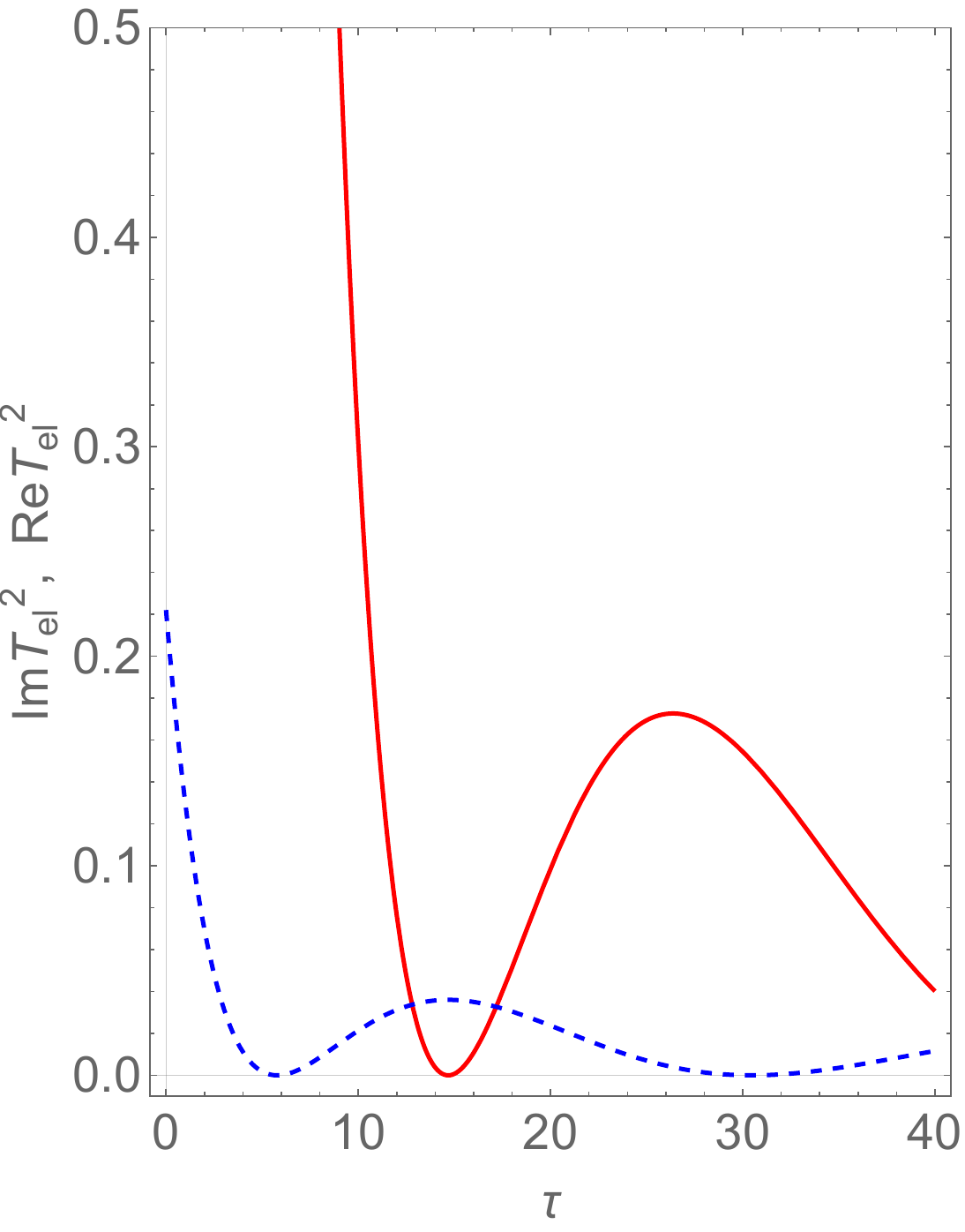}
\vspace{-0.2cm} \caption{Left panel: function $\Phi(\tau)$ from Eq.~(\ref{eq:PhiHD2}) -- solid red line,
and $d(\tau \Phi(\tau))/d\tau$ -- dashed blue. Right panel: Contributions of imaginary (solid red)
and real (dashed blue) parts of the scattering amplitude to the elastic cross section. For the real
part the $\rho$ parameter is 0.15. }%
\label{fig:Phi}%
\end{figure}

It is instructive to illustrate this in the black disc approximation, where
\begin{equation}
\Phi(\tau)={2 \pi}\frac{J_{1}(\sqrt{\tau})}{\sqrt{\tau}},
\label{eq:PhiHD2}
\end{equation}
which is plotted in the left panel of Fig.~\ref{fig:Phi}. In the right panel we plot $\Phi^2$,
which is responsible for the imaginary part contribution to the cross section.
Therefore we have
\begin{equation}
\Phi(\tau_{\mathrm{dip}})=0,~~\frac{d}{d\tau}\Phi(\tau)|_{\mathrm{bump}%
}=0.
\label{eq:dipbumpconds}%
\end{equation}

To include the real part of $T_{\mathrm{el}}$ we have to impose crossing
symmetry~\cite{DiasdeDeus:1975ybq} on (\ref{eq:ansatz0}). In the Regge limit (all masses equal zero,
$s\gg-t$) the  crossing relation takes the following form%
\begin{equation}
\tilde{T}_{\text{el}}(u,t)\simeq\tilde{T}_{\text{el}}(-s,t)=\tilde
{T}_{\text{el}}^{\ast}(s,t). \label{eq:crossing}%
\end{equation}
Therefore, following \cite{DiasdeDeus:1975ybq,DiasdeDeus:1977af} we assume%
\begin{equation}
\tilde{T}_{\text{el}}(s,\tau)=isR^{2}(-is)\Phi\Big( \left\vert t\right\vert
R^{2}(-is)\Big) ,
\end{equation}
which satisfies (\ref{eq:crossing}). We will now identify real and imaginary
parts of the amplitude observing that%
\begin{equation}
-is=e^{y-i\pi/2}.
\end{equation}
Expressing $R^{2}$ in terms of rapidity $y=\ln s$ rather than $s$, and
expanding, we get \cite{DiasdeDeus:1975ybq}%
\begin{equation}
R^{2}(-is) \rightarrow R^{2}\left(  y-i\frac{\pi}{2}\right)  \simeq
R^{2}(y)-i\frac{\pi}{2}\frac{dR^{2}(y)}{dy} \, .\label{eq:expansionR}%
\end{equation}
Expanding further we arrive at%
\begin{equation}
\Phi\Big( \left\vert t\right\vert R^{2}(-is)\Big) \simeq\Phi\left(
\tau\right)  -i\frac{\pi}{2}\frac{d\Phi\left(  \tau\right) }{d\tau}%
\frac{dR^{2}(y)}{dy}\left\vert t\right\vert . \label{eq:expansion}%
\end{equation}
Keeping only linear terms from expansions (\ref{eq:expansionR}) and
(\ref{eq:expansion}) we get~\cite{DiasdeDeus:1975ybq,DiasdeDeus:1977af}%
\begin{align}
\operatorname{Im}\tilde{T}_{\text{el}}(s,\tau)  &  = sR^{2}(y)\Phi\left(
\tau\right) ,\nonumber\\
\operatorname{Re}\tilde{T}_{\text{el}}(s,\tau) &  =s\frac{\pi}{2}\frac
{dR^{2}(y)}{dy}\frac{d}{d\tau}\left(  \tau\Phi\left( \tau\right)  \right)  \,
.\label{eq:ImRe}%
\end{align}
In Fig.~\ref{fig:Phi} we plot both $\Phi(\tau)$ and $d(\tau \Phi(\tau))/d\tau$ for the hard disc
toy model (\ref{eq:PhiHD2}), and -- in the right panel -- the corresponding contributions to the cross section.
We see that the cross section is dominated by the imaginary part squared, except for the dip, where the
real part dominates (for illustration purposes we have assumed $\rho=0.15$).

\begin{figure}[h]
\centering
\includegraphics[width=9.cm]{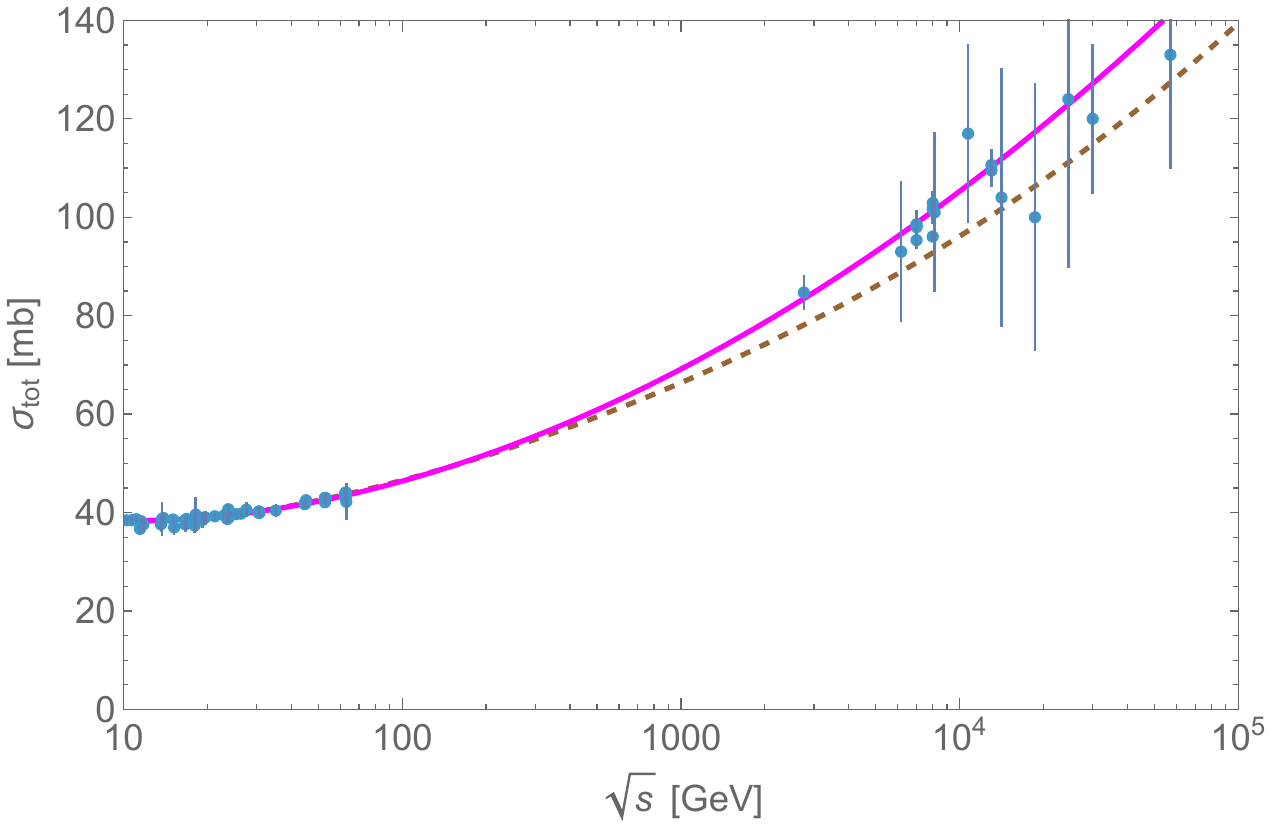} \vspace{-0.2cm} \caption{Total $pp$
cross-section in mb as a function of $\sqrt{s}$ in GeV. Points at small
$\sqrt{s} < 100$~GeV are from the ISR, points above 1~TeV are from the LHC
(small error bars) and from cosmic rays. Solid magenta line corresponds to the
PDG parametrization (\ref{eq:parPDG}) and brown dashed line to (\ref{eq:parDL}%
).}%
\label{fig:total}%
\end{figure}

\section{Comparison with the data}

Formulas (\ref{eq:ImRe}) result in an absolute prediction for the $\rho$
parameter
\begin{equation}
\rho=\frac{\operatorname{Re}\tilde{T}_{\text{el}}(s,\tau=0)}{\operatorname{Im}%
\tilde{T}_{\text{el}}(s,\tau=0)} = \frac{\pi}{2 }\frac{1}{R^{2}(y)}%
\frac{dR^{2}(y)}{dy} \, .
\label{eq:rhopred}%
\end{equation}
Relation (\ref{eq:rhopred}) has been derived, at least
parametrically, from the dispersion relations long time ago (see {e.g.}
Ref.~\cite{Ryskin:2024qpq}). However, it is instructive to compare it with the
current data. For illustration purposes we use two analytic parametrizations
of the total $pp$ cross section. The first one is the COMPETE parametrization
\cite{Cudell:2001pn} quoted in PDG~2010~\cite{PDG2010}
\begin{equation}
\sigma_{\mathrm{tot}}^{\mathrm{PDG}}(s)=Z+C \ln^{2}\left(  \frac{s}{s_{0}%
}\right)  +Y_{1}\left(  \frac{s}{s_{1}}\right)  ^{-\eta_{1}}-Y_{2}\left(
\frac{s}{s_{1}}\right)  ^{-\eta_{2}} \, ,
\label{eq:parPDG}%
\end{equation}
where $Z=35.45$~mb, $C=0.308$~mb, $Y_{1}=42.53$~mb, $Y_{2}=33.34$~mb,
$s_{0}=28.94$~GeV$^{2}$, $s_{1}=1$~GeV$^{2}$ and $\eta_{1}=0.458$, $\eta
_{2}=0.545$. The second one is from Ref.~\cite{Donnachie:1992ny} by Donnachie
and Landshoff
\begin{equation}
\sigma_{\mathrm{tot}}^{\mathrm{DL}}(s)= A \left(  \frac{s}{s_{1}}\right)
^{\alpha}+B\left(  \frac{s}{s_{1}}\right)  ^{\beta} \ ,
\label{eq:parDL}%
\end{equation}
where $A=21.70$~mb, $B=56.08$~mb, $\alpha=0.0808$, $\beta=-0.4525$ and
$s_{1}=1$~GeV$^{2}$. They are plotted in Fig.~\ref{fig:total} together with
data points (see the web page of PDG~2022~\cite{PDG2022}). One can see that the
PDG parametrization (\ref{eq:parPDG})  covers the whole energy range from the ISR to the LHC (including the
$p\bar{p}$ data not shown on the plot),
whereas the older parametrization by Donnachie and Landshoff (\ref{eq:parDL}) devised to describe
low energy data undershoots the LHC points. 

\begin{figure}[h]
\centering
\includegraphics[width=9.cm]{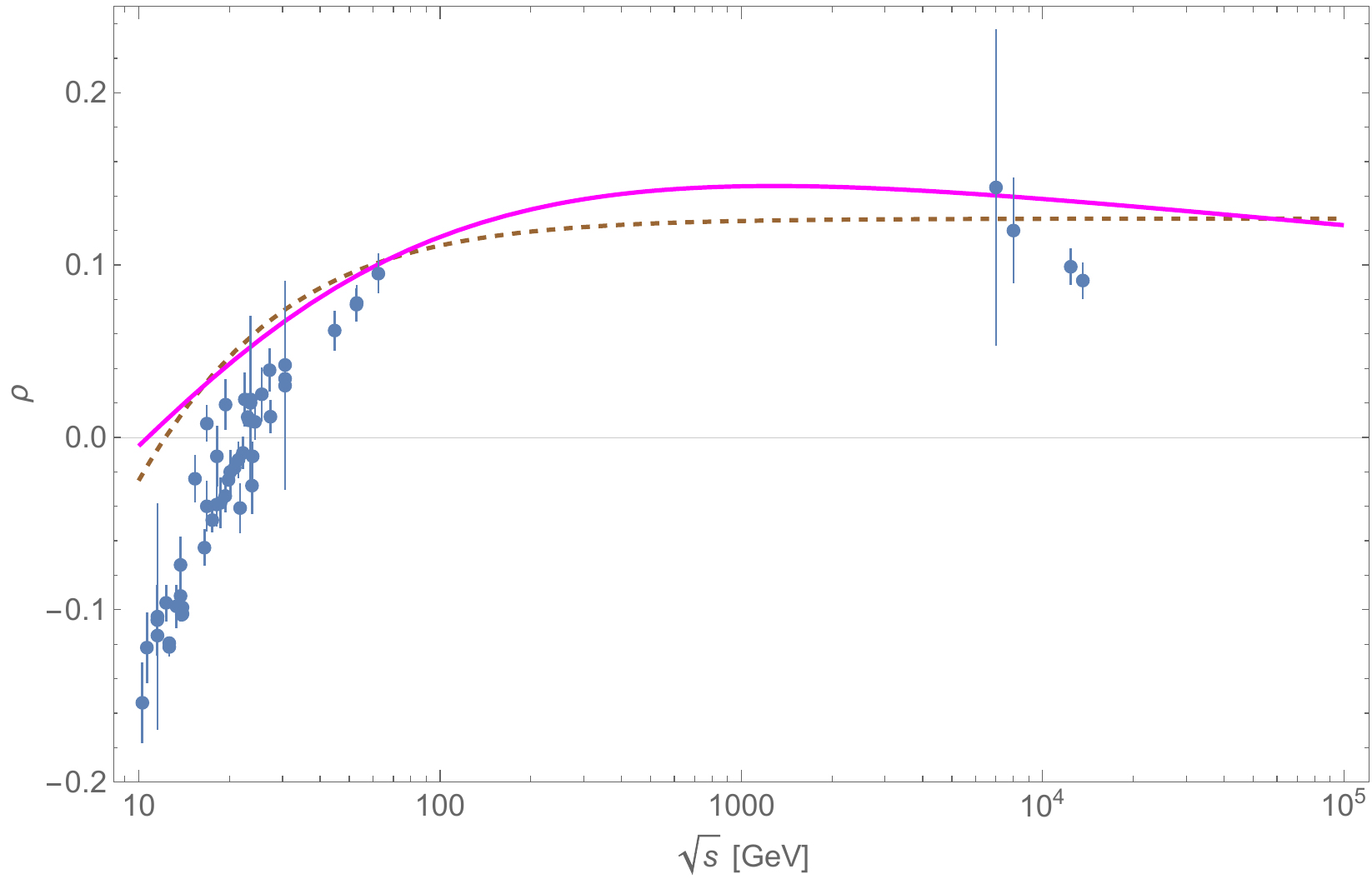} \vspace{-0.2cm}
\caption{Parameter $\rho$ (\ref{eq:rhopred})
as a function of $\sqrt{s}$ in GeV. Points at small
$\sqrt{s} < 100$~GeV are from the ISR, points above 1~TeV are from the LHC.
 Solid magenta line corresponds to the
PDG parametrization (\ref{eq:parPDG}) and brown dashed line to (\ref{eq:parDL}).}%
\label{fig:rho}%
\end{figure}

In Fig.~\ref{fig:rho} we plot the parameter $\rho$ computed according to Eq.~(\ref{eq:rhopred}) together with
the data \cite{PDG2022}. We can see that both the PDG parametrization (\ref{eq:parPDG})
and the DL one  (\ref{eq:parDL}) describe the $\rho$ parameter rather well, overshooting the lower 
energy ISR points. This is not a surprise since expansion (\ref{eq:expansion}) is in fact asymptotic.
Note that both normalization and the energy dependence are reproduced without any adjustable parameter.
The last two points in Fig.~\ref{fig:rho} correspond to
two different estimates by TOTEM \cite{TOTEM:2017sdy}. The fact that they are 
lower than our predictions (and, in fact, lower than in many other more precise models and parametrizations) 
has led to the conjecture that this deviation is a sign of the odderon~\cite{Csorgo:2019ewn,TOTEM:2020zzr,Ryskin:2024qpq,Martynov:2017zjz} 
(see however  \cite{Khoze:2017swe,Khoze:2018kna}).
If so, our analysis should be 
extended to include the  $C$ parity odd amplitude, which is beyond the scope of the present study.

Now, from Eq.~(\ref{eq:dipbumpconds}) we can compute ratio
$\mathcal{R}_{\mathrm{bd}}(s)$ (\ref{eq:Rbd}). Indeed, we have
\begin{align}
\Phi(\tau_{\mathrm{dip}})=0 \rightarrow &  \operatorname{Im}\tilde
{T}_{\text{el}}(s,\tau_{\mathrm{dip}})=0 \, ,
\nonumber\\
&  \operatorname{Re}\tilde{T}_{\text{el}}(s,\tau_{\mathrm{dip}})=s\frac{\pi
}{2}\frac{dR^{2}(y)}{dy}\frac{d}{d\tau} \Phi(\tau_{\mathrm{dip}}) \, ,
\end{align}
and
\begin{align}
\frac{d}{d\tau} \Phi(\tau_{\mathrm{bump}}) =0 \rightarrow &  \operatorname{Im}%
\tilde{T}_{\text{el}}(s,\tau_{\mathrm{dip}})= sR_{1}^{2}(y)\Phi(\tau
_{\mathrm{bump}}) \, ,
\nonumber\\
&  \operatorname{Re}\tilde{T}_{\text{el}}(s,\tau_{\mathrm{dip}})=s\frac{\pi
}{2}\frac{dR^{2}(y)}{dy} \Phi(\tau_{\mathrm{bump}}) \, .
\end{align}
Therefore%
\begin{align}
\left.  \frac{d\sigma}{dt}\right\vert _{\text{dip}}  &  
=\frac{R^{4}(y)}{4\pi} \, \rho^{2}(y) \, \left(  \tau_{\text{dip}}\frac{d}{d\tau}
\Phi (\tau_{\mathrm{dip}})\right)^{2}%
\end{align}
and%
\begin{align}
\left.  \frac{d\sigma}{dt}\right\vert _{\text{bump}}  &  
 =\frac{R^{4}(y)}{4\pi} \, \Big(  1+\rho^{2}\left(  y\right)  \Big)\,
\Phi^{2}(\tau_{\mathrm{bump}}) \, .
\end{align}%
Hence, we have a one parameter prediction for the ratio
\begin{equation}
{\cal R}_{\rm bd}(s)=\frac{d\sigma/dt(t_{\text{bump}})}{d\sigma/dt(t_{\text{dip}})}=c_{0}%
\frac{1+\rho^{2}\left(  y\right)  }{\rho^{2}\left(  y\right)  },
\label{eq:ratioRbd}
\end{equation}
where constant $c_0$ is given in terms of  function $\Phi$%
\begin{equation}
c_{0}=\frac{\Phi^{2}(\tau_{\mathrm{bump}})}{\left(  \tau_{\text{dip}}\frac
{d}{d\tau}\Phi (\tau_{\mathrm{dip}}) \right)  ^{2}} \, .
\label{eq:Rbdc0}
\end{equation}

\begin{figure}[h]
\centering
\includegraphics[width=9.cm]{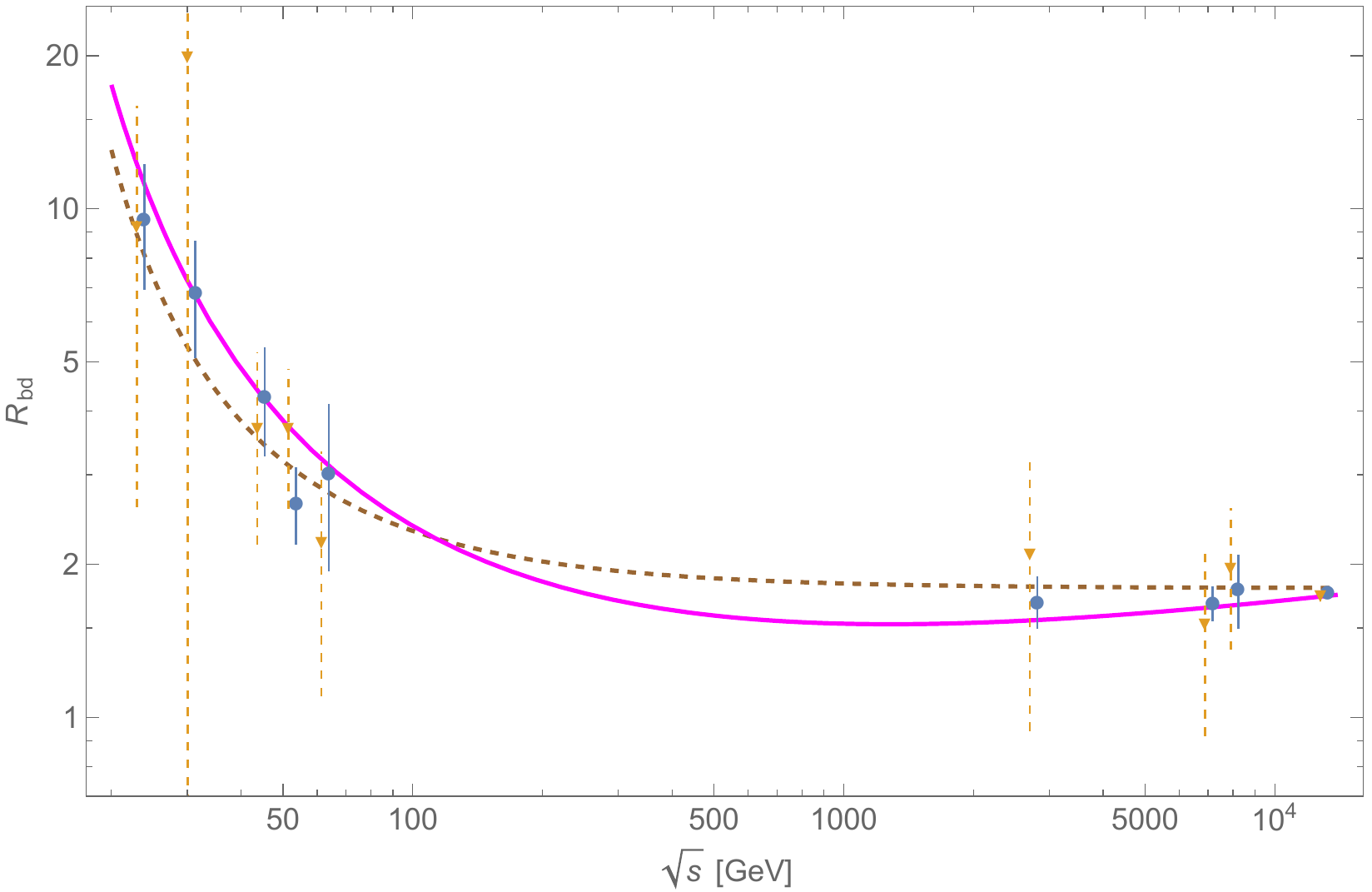} \vspace{-0.2cm}\caption{Ratio
${\cal R}_{\mathrm{bd}}$ as a function of $\sqrt{s}$ in GeV. Data: 
brown triangles from Ref.~\cite{Baldenegro:2024vgg},
blue circles from Ref.~\cite{TOTEM:2020zzr}. Theoretical curves (\ref{eq:ratioRbd})
as in Figs.~\ref{fig:total} and \ref{fig:rho}.}%
\label{fig:Rbd}%
\end{figure}

In Fig.~\ref{fig:Rbd} we plot ratio (\ref{eq:ratioRbd}) for two parametrizations (\ref{eq:parPDG}) and (\ref{eq:parDL})
with $c_0=0.012 \div 0.013$. One can see that the ratio data are well reproduced by both parametrizations.

And finally we can compute the energy dependence of the total elastic cross section%
\begin{align}
\sigma_{\text{el}}(s)  & 
 =\frac{1}{4\pi R^{2}(y)}\left[  {R^{4}(y)}%
{\int}d\tau{\Phi}^{2}{(\tau)+}\left(  {\frac{\pi}{2}\frac{dR^{2}(y)}{dy}}\right)^{2}%
{\int}d\tau\left(  {\frac{d}{d\tau}\left(  \tau\Phi (\tau)\right)  }\right)^{2}\right] \notag \\
 &=\frac{R^{2}(y)}{4\pi}\left(  1+c_{1}\rho^{2}(y)\right)  \times%
{\displaystyle\int}
d\tau{\Phi}^{2}{(\tau)}\, ,%
\end{align}
where%
\begin{equation}
c_{1}=\frac{%
{\displaystyle\int}
d\tau\left(  {\frac{d}{d\tau}\left(  \tau\Phi (\tau)\right)  }\right)
^{2}}{%
{\displaystyle\int}
d\tau{\Phi}^{2}{(\tau)}}.
\label{eq:sigeltot}
\end{equation}

We see from Eq.~(\ref{eq:sigeltot}) that the energy dependence of the total elastic
$pp$ cross section is modified with respect to the total cross section
by a factor
\begin{equation}
\frac{\sigma_{\text{el}}(s)}{\sigma_{\text{tot}}(s)}\sim\left(  1+c_1 \rho
^{2}(s)\right) \, .
\end{equation}
At the ISR the $\rho$ parameter is very small and, despite the fact that it rapidly rises with energy, its influence
on  $\sigma_{\rm el}$ is negligible.\footnote{To be more quantitative we need to compute $c_1$.} 
At the LHC the $\rho$ parameter is slightly larger but it is almost constant
(see Fig.~\ref{fig:rho}) and therefore does not modify the energy dependence of  $\sigma_{\rm el}$ as well.
This result is of course in conflict with the data. However, note that  formula (\ref{eq:sigeltot}) assumes that GS 
is valid {\sl everywhere} in $t$. This is certainly not true outside
the dip\,-\,bump region, although at the ISR GS surprisingly holds gown to a very small $t$ \cite{Baldenegro:2024vgg}. 
Therefore, the inability
of GS to describe the energy dependence of the total elastic cross section at the LHC should be attributed to
the GS violation outside the dip\,-\,bump region, most importantly at small $t$. 

\section{Summary and discussion}

To summarize: we have shown that properties of the dip\,-\,bump structures of the differential  elastic cross sections
can be interpreted assuming geometric scaling both at the ISR and at the LHC. The scaling variable is $\tau \sim \sigma_{\rm tot}(s) |t|$.
Assuming GS and crossing, and applying
 expansion (\ref{eq:expansionR}) and (\ref{eq:expansion}),
we have been able to identify imaginary and real parts of the scattering amplitude and derive formulas for the parameter $\rho$
and the ratio ${\cal R}_{\rm b d}$, which reproduce the data,
despite the fact that the analysis presented here, which extends the approach of  
Refs.~\cite{DiasDeDeus:1973lde,Buras:1973km,DiasdeDeus:1975ybq,DiasdeDeus:1977af} to the LHC,
is only of qualitative character. Therefore, one should not expect too high accuracy when comparing with the data.
The energy dependence of the total elastic $pp$ cross section 
is, however, not reproduced. We have attributed this to the GS violation outside  the dip\,-\,bump region, at small values of $t$,
which dominate the total elastic cross section.
Indeed, it is well known, that the
phenomenological formula 
\begin{equation}
\frac{\sigma_{\text{el}}(s)}{\sigma_{\text{tot}}(s)}\sim \frac{\sigma_{\rm tot}(s)}{B(s)} \left(  1+ \rho
^{2}(s)\right) \, ,
\label{eq:et2tot}
\end{equation}
where $B(s)$ is a slope of the diffractive peak, works very well to an accuracy of a few percent (see e.g. \cite{Block:2006hy,Broniowski:2018xbg}).
In deriving the formula (\ref{eq:et2tot}) one basically neglects the dip\,-\,bump structure taking into account only the diffraction peak,
which gives the dominant contribution to the total elastic cross section, in a form which does not exhibit GS~\cite{Block:2006hy}.

At this point it is interesting to compare our results with the so called $H$ scaling proposed in Ref.~\cite{Csorgo:2019ewn}, where function
$H(x)$ defined as
\begin{equation}
H(x)=\frac{1}{B(s)\sigma_{\rm el}(s)} \frac{d\sigma_{\rm el}}{dt}(s,t)
\label{eq:Hscaling}
\end{equation}
should depend o the scaling variable $x=B(s)|t|$ alone. At small $t$ we can neglect the real part of the scattering amplitude where, if GS holds,
the cross section reads
\begin{equation}
\frac{d\sigma_{\rm el}}{dt}(s,t)=\frac{\sigma_{\rm tot}^2(s)}{4 \pi} \Phi^2(\tau),
\end{equation}
and
\begin{equation}
H(x)=\frac{1}{4 \pi} \frac{\sigma_{\rm tot}^2(s)}{B(s)\sigma_{\rm el}(s)} \Phi^2(\tau).
\end{equation}
However, asymptotically $\sigma_{\rm tot}(s)/\sigma_{\rm el}(s) \rightarrow {\rm const.}$ and  $\sigma_{\rm tot}(s)/B(s) \rightarrow {\rm const.}$ \cite{Block:2006hy},
hence $x \rightarrow \tau$ and the $H$ scaling and GS coincide. The difference in sub-asymptotic energies is of course important for phenomenology. To account
for a difference between $B(s)$ and $\sigma_{\rm tot}(s)$ one could use the property known as the stationary point, which provides a relation
between these two quantities, which is valid from the ISR to the LHC~\cite{Samokhin:2017kde}.

The above discussion shows that in order to improve phenomenology of GS one needs to include GS violations at small $t$.
It would be interesting to extend the present study to the $p\bar{p}$ case, and to include the odderon amplitude. One may also try to find corrections
to the expansions (\ref{eq:expansionR}) and (\ref{eq:expansion}).

Finally, the question arises, what is the physical origin of GS and why it extends over such a wide energy range. GS is surprising because
it is not a general prediction of the phenomenological models. Indeed, early models based on unitarity \cite{Gotsman:1993vd} or
multi pomeron exchanges \cite{Dubovikov:1976kq} lead to the opacity function, which has additional energy dependence, which 
violates GS: $\Omega(s,b)=\nu(s) F(b/R(s))$, unless $\nu(s)$ is tuned to be energy independent (or weakly dependent) in some limitted energy
range. Typically, for Gaussian $F$ one recovers GS at asymptotically large energies \cite{Gotsman:1993vd,Dubovikov:1976kq}.
Therefore, it was believed that at finite energies GS can arise only accidentally \cite{Barger:1974vg}. The present analysis seems to contradict such claims.
Nevertheless, some GS violation should be incorporated in the amplitude parametrization (\ref{eq:ansatz0}). Such an attempt was made, 
for example, in  Ref.~\cite{Kawasaki:1995rx}.

The opacity function in Gaussian form is useful because it allows the analytical calculation of $\sigma_{\rm tot}$ and/or $\sigma_{\rm el}$.
However, its Fourier transform results in an unrealistic amplitude with too dense distribution of dips and bumps. Grey disc parameterizations
 provide a much better description of the differential elastic cross-section \cite{Kamal:1975}. Therefore, understanding the shape of of the
 imaginary part of the elastic amplitude in $b$ space can provide insight into the physical nature of GS. 
 
 Another possibility is discussed in Ref.~\cite{Peschanski:2024tlr}, where the authors try to link the scaling at the LHC to the 
 hard elastic scattering and deep inelastic $ep$ scattering. In tis case hard probes appearing in $pp$ scattering
 are identified with the so-called {\sl hot spots} defined as small regions of high gluon density in an interacting particle at high
energy. However, it remains an open question whether the ISR energy can be considered high enough in this case.

\section*{Acknowledgments}
The author acknowledges CERN TH Department for hospitality while this research was being carried out,
and would like to thank T.~Cs\"org\H{o}, Ch.~Royon and A.~Staśto for discussions and remarks.

\end{document}